# Considerations in the Use of Machine Learning Force Fields for Free Energy Calculations

*Orlando A. Mendible Barreto, Jonathan K. Whitmer, and Yamil J. Colón**

Department of Chemical and Biomolecular Engineering, University of Notre Dame, Notre Dame, IN, 46556. *Corresponding author, email: ycolon@nd.edu

***ABSTRACT:***

Machine learning force fields (MLFFs) promise to accurately describe the potential energy surface of molecules at the ab initio level of theory with improved computational efficiency. Within MLFFs, equivariant graph neural networks (EQNNs) have shown great promise in accuracy and performance and are the focus of this work. The capability of EQNNs to recover free energy surfaces (FES) remains to be thoroughly investigated. In this work, we investigate the impact of collective variables (CVs) distribution within the training data on the accuracy of EQNNs predicting the FES of butane and alanine dipeptide (ADP). A generalizable workflow is presented in which training configurations are generated with classical molecular dynamics simulations, and energies and forces are obtained with *ab initio* calculations. We evaluate how bond and angle constraints in the training data influence the accuracy of EQNN force fields in reproducing the FES of the molecules at both classical and *ab initio* levels of theory. Results indicate that the model's accuracy is unaffected by the distribution of sampled CVs during training, given that the training data includes configurations from characteristic regions of the system's FES. However, when the training data is obtained from classical simulations, the EQNN struggles to extrapolate the free energy for configurations with high free energy. In contrast, models trained with the same configurations on ab initio data show improved extrapolation accuracy. The findings underscore the difficulties in creating a comprehensive training dataset for EQNNs to predict FESs and highlight the importance of prior knowledge of the system's FES.

# 1 Introduction

Machine learning forcefields (MLFFs) have emerged as a new tool that aims to maintain the accuracy of *ab initio* calculations while reducing the computational cost by orders of magnitude and maintaining a linear scaling with the number of atoms in the system.[1–4] Many MLFF models have been developed and used for new materials,[5–8] chemical reactions,[9–11] large-scale simulations,[12–14] and understanding complex interatomic and intermolecular interactions[5,15,16]. The main differences between models are the descriptors used to describe the local atomic environments within the system, such as the Smooth Overlap of Atomic Positions (SOAP)[17], the Atomic Cluster Expansion (ACE)[18], among others[19], and the type of machine learning methods, mainly neural networks[20] or kernel-based[21,22] methods. Various combinations of approaches and atomic representation methods have given rise to an overwhelming amount of open-source computational models to train MLFFs, including DeepMD-Kit[23], SchNet[24], TorchMD[25], sGDLS[26], NEQUIP[27], and many more[28]. Among these approaches, equivariant graph neural networks (EGNN) have shown optimal accuracy and scalability[29]. Specifically, the Allegro[30] EGNN, which will be the focus of this study, represents a cutting-edge approach to learning interatomic potentials through deep neural networks. Unlike other methods relying on atom-centered message passing, Allegro employs strictly local equivariant representations, enabling it to capture many-body interactions with greater accuracy and scalability. This distinction is evidenced by its exceptional accuracy and performance in simulating systems with up to 100 million atoms[31] and achieving high accuracy across a range of materials[30,32,33].

Limited literature is currently available where EGNN is tested for predicting the free energy surface (FES) of molecular systems using enhanced sampling molecular simulations. Ple et al.[34] employed the FENNIX (Force-field-Enhanced Neural Network Interactions) model, which uses Allegro's representations of atomic pairs and then passes this data through an additional neural net responsible for predicting the property of interest. They trained a FENNIX model for alanine dipeptide (ADP) in explicit water using data from classical MD and qualitatively recovered the FES but underestimated high free energy barriers. Similarly, Kovács et al.[35] used the MACE-OFF23 model to predict the various properties of a set of organic molecules. However, the predictions made from this message-passing neural network model for the FES of alanine-tripeptide in explicit solvent demonstrate the model's ability to predict the location of free energy minima but showed limitations with high free energy configurations. More recently, Pérez-Lemus et al.[36] used Allegro to recover the FES of ADP in a vacuum and explicit water with good accuracy and mentioned potential limitations of including bond and angle constraints in the training data. In this study, we provide an in-depth look into the effects of including these constraints in the training data on the accuracy of FES predictions. In a different publication, they also used EGNNs to determine the FES of methane activation over a Ni surface, providing insights into the dynamic interplay between the molecule and the catalytic material[37].

As systems become more complex, including diverse configurations in the training data becomes challenging. Efforts have been made to enhance configurational sampling in training data through methods like FLARE[38] and DP-GEN[39], which automates data preparation and model training using concurrent and active learning techniques. However, these approaches may still lack configurations representative of the underlying FES due to dependence solely on model deviations and uncertainties. Moreover, recent results highlight the ability of equivariant models to recover the FES of the system even when not all the configurations are included in the training data[36].

In this work, we study the data size and collective variable (CV) distribution effects on the accuracy of Allegro to determine the FES of butane and ADP molecules in a vacuum using metadynamics simulations. The rotation of butane's methyl groups around the central carbon-carbon bond gives rise to various conformers mapped by the torsional angle of its carbon atoms[40]. Additionally, its FES has been extensively studied, providing a valuable reference to compare results[41–43]. Similarly, the configurational diversity of ADP stems from the rotation of its phi (ϕ) and psi (ψ) dihedral angles, which succinctly represent the molecule's configurational space[44,45–47], and the FES has also been extensively studied[48–50]. For these reasons, butane and ADP's FES are ideal candidates for examining the interplay between structural transitions and energetic stability. Based on previous efforts to study the effect of including bond and angle constraints in the training data on the accuracy of MLFFs, we also study the impact of these constraints specifically on the accuracy of free energy predictions on two types of models, one with data from classical molecular dynamics and the other from *ab initio* calculations. The information obtained is valuable for understanding why analyzing CVs and potential energy distribution in your system when generating training data is essential. It also highlights considerations to remember when training an EGNN to predict the free energy landscape of molecular systems.

# 2 Methods

## 2.1 Computational methods

Simulations at the classical level of theory were conducted using LAMMPS[51], and interatomic interactions of butane and ADP were modeled using the OPLS-AA[52] and AMBERSB99[53,54] forcefields, respectively. Classical simulations were conducted using a timestep of 1 fs, a cutoff of 12 Å, and Coulombic interactions were computed using the particle-particle/particle-mesh (PPPM) method with a tolerance of $1 \times 10^{-6}$.

*Ab initio* calculations, including molecular dynamics and single-point calculations, were performed using the Quickstep module in CP2K[55]. The pseudopotentials and basis sets are taken from Niemöller et al.[56], and they include the use of the revPBE[57] functional and corresponding PBE Goedecker–Teter–Hutter pseudopotentials for core electrons[58–60], in combination with the molecularly optimized basis set (MOLOPT-DZVP-GTH) [61]. CUTOFF density was set to 800 Ry and a relative CUTOFF of 60 Ry. Dispersion interactions were accounted for by applying the DFT-D3 correction[62,63]. The self-consistent field (SCF) was set up with an accuracy threshold of $10^{-6}$. The DIIS[64] (direct inversion in the iterative subspace) minimizer and FULL_SINGLE_INVERSE preconditioner were applied.

## 2.2 Reference free energy surfaces (FES)

The reference FES of the systems of interest were determined to facilitate the assessment of the extent to which the training data samples the FES. At the classical level of theory, the FES of butane was determined using a metadynamics simulation of 10 ns to bias the rotation of the phi dihedral. A similar simulation was conducted for ADP, biasing the more complex molecule's phi and psi dihedral angles. These simulations were performed using LAMMPS and PLUMED[65] code, and the biasing parameters include the addition of Gaussians every 500 steps, with a height of 0.239 kcal/mol, a width of 0.3 radians, and a bias factor of 4.

At the *ab initio* level of theory, the reference FES of butane was obtained by running a 700 ps metadynamics simulation at the *ab initio* level of theory. This was achieved using CP2K coupled with PLUMED and the pseudopotentials and basis sets described in the previous sections. For this simulation, the phi dihedral of butane was biased by adding Gaussians every 100 steps, with a height of 0.239 kcal/mol, width of 0.3 radians, and a bias factor of 4. Due to ADP's complex free energy landscape, the reference FES at the *ab inito* level of theory was obtained only as a function of its phi dihedral. This was achieved by running an umbrella sampling simulation in which twenty walkers were used to run 50 ps restrained simulations at CV values between $-\pi$ and $\pi$. The total simulation time is one ns. Afterward, the WHAM[66] method was applied to recover the unbiased FES of the molecule.

## 2.3 Training and test data preparation

The training data was generated by producing configurations with classical MD, performing single-point calculations (SPC) on them at the *ab initio* level of theory, and storing the coordinates, atom types, and the resulting energies and forces.

A simulation of 10 ns at 350 K was sufficient to obtain enough configurations to sample all the regions in the system's FES. Butane configurations located at specific areas of the free energy landscape were selected, and their coordinates, potential energy, and atomic forces were extorted and stored as the training data sets from classical MD. The coordinates from these data sets were used as inputs for SPC calculations, which determined energies and forces at the *ab initio* level of theory. Finally, the resulting values were stored in the SPC training data sets. The same procedure was performed again using independent simulations to generate the test data sets. Following this approach, four training/test data sets were generated, each sampling the FES of the system in a distinct way. A histogram of the distribution of the CVs in the training and test data is plotted with the reference FES of butane at the *ab initio* level of theory in Figure 1.

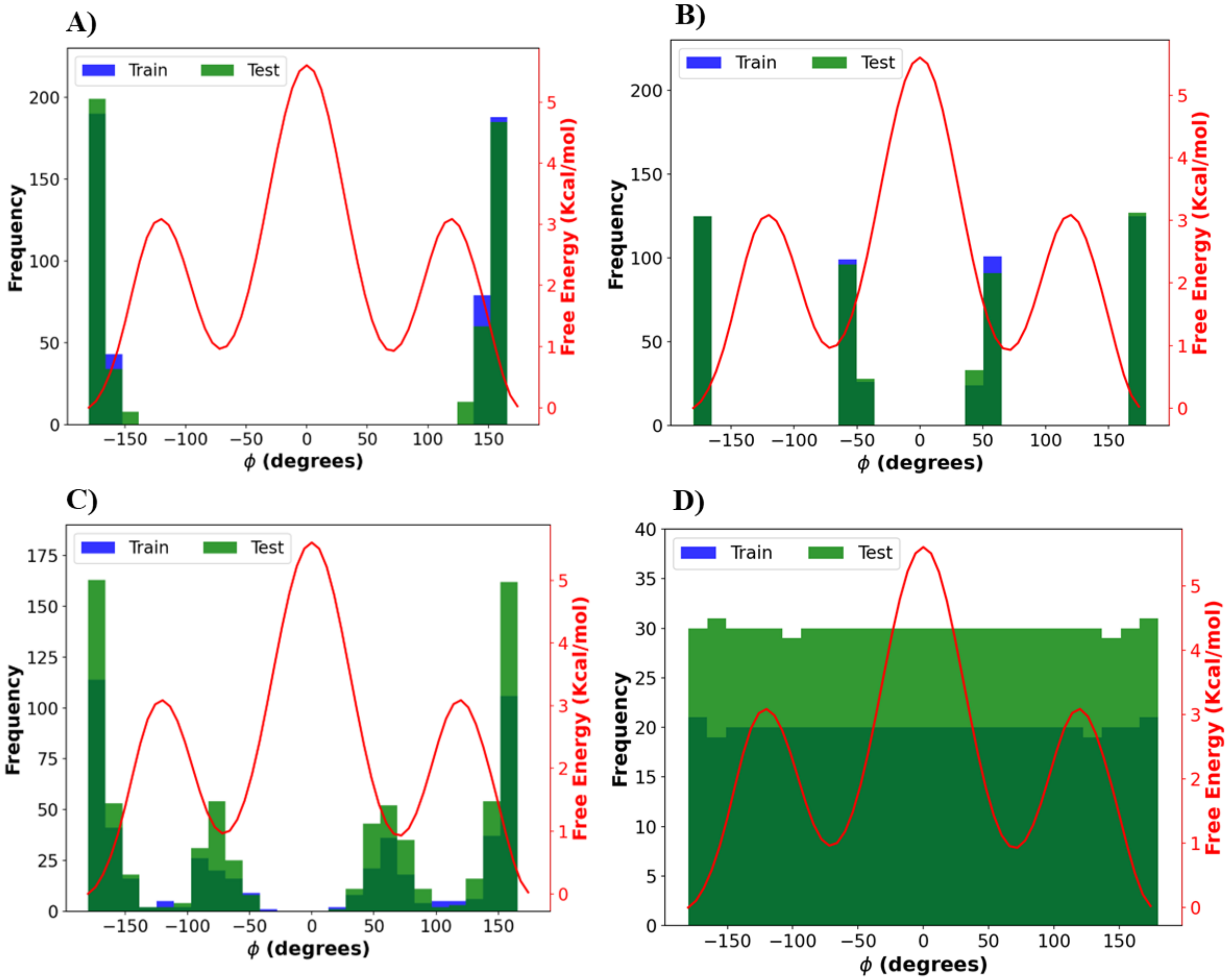

*Figure 1. Histogram of butane CVs included in A) Global minimum distribution (GM), B) Local minimum distribution (LM), C) Unbias distribution (Unb), and D) Uniform distribution (Uni) training/test dataset over butane's FES.*

Each dataset had 500 configurations, and their names highlight the region of the FES covered by the samples in each dataset. Therefore, they are named *Global minimum distribution* (GM), *Local minimum distribution* (LM), *Unbiased distribution* (Unb), and *Uniform distribution* (Uni). The GitHub repository has all the inputs and code to generate the datasets.

For ADP, two different methods were used to generate configurations of interest. First, a 20 ns unbiased simulation was performed at 350 K to obtain configurations that represented the system's free energy basins. From the generated trajectory, we selected 2500 samples to create the unbiased distribution data set (A-UD). Then, we defined the phi and psi values representing four free energy minima and selected the 2500 configurations with CV values closer to the defined minimums. Configurations were stored in the Only Minima distribution (A-OM) datasets.

The second approach includes the generation of ADP configurations with specific CV values by implementing restrained simulations. A two-dimensional grid is generated to uniformly represent the combinations of phi and psi values across the system FES. The CV values of each point were used as the fixed value, or reference point, for restrained simulations at 500 K with a harmonic force constant of 250 kJ/mol. The generated trajectories include various configurations with CV values near the defined reference point, but only the configuration in which CV was closest to the reference point was selected to be part of the training dataset. The unbiased energy of the chosen configurations was calculated by running a single-step simulation using a time step of 0.01 fs, and the resulting energies and atomic forces were extracted and stored for training the models at the classical level of theory. Finally, the coordinates of this dataset were used as inputs for SPCs from which the energy and atomic forces were extracted to generate the dataset at this level of theory. This approach was followed to create the uniform distribution (A-UD) and the characteristic regions (A-CRD) distributions. The CV values of each ADP configuration are plotted on top of the FES of ADP at the classical level of theory in Figure 2. This plot facilitates the analysis of configurations covering specific regions of the system's FES. The required input files, code, and additional details are available in this project's GitHub repository.

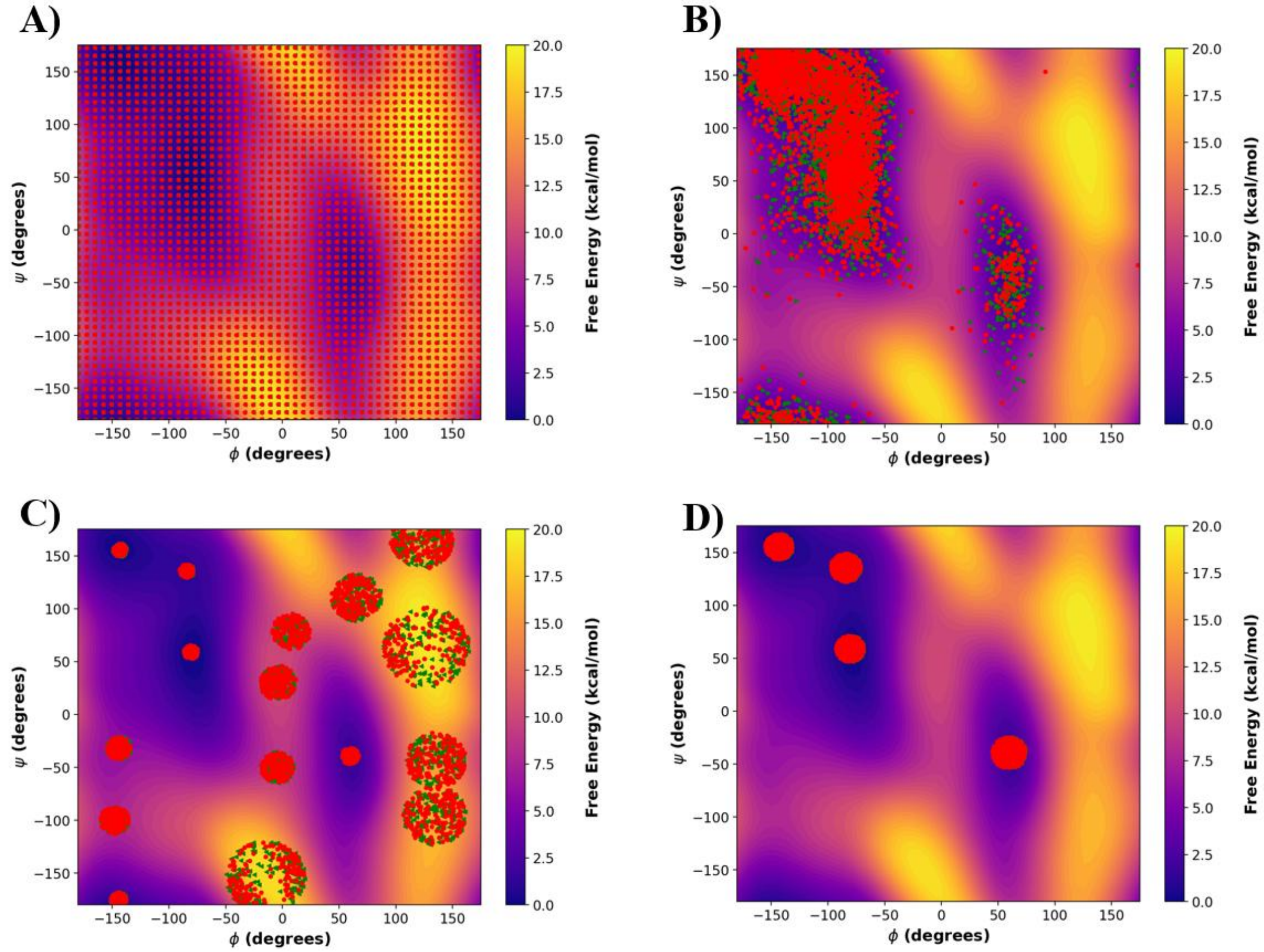

*Figure 2. Two-dimensional Histogram of ADP CVs included in each training/test dataset over its 2D-FES. All distributions have 2500 configurations. In A) the Uniform distribution (A-Uni), in B) the Unbias distribution (A-Unb), in C) the Characteristic regions distribution (A-CR), and in D) the Only Minimums distribution (A-OM).*

## 2.4 Equivariant Interatomic Potential

The initial model was selected with the suggested hyperparameters found in the examples of Allegro[67] configurations. Hyperparameter optimization was conducted using the Weight-and-Biases software[68]. More information can be found in Section I of the Supplemental Information (SI). For all the training data sets, 80% of the data was used for training, and 20% was used for validation. The training process incorporated a local cutoff of 4.0 Å, and a polynomial envelope function with a parameter value of p=6 was employed for the Bessel functions. This function facilitated the decay of interactions as the interatomic distance increased. A maximum rotational number of lmax=3 was also set for spherical harmonics. Three layers were used, and for each, the number of features was 32; the two-body latent dimension of the neural network architecture comprised five layers with dimensions [64, 128, 256, 512, 1024] and a latent multi-layer perceptron with dimension [1024,1024,1024]. SiLu nonlinearities were applied to the outputs of each hidden layer. Subsequently, the latent Multi-Layer Perceptron (MLP) consisted of three layers, each with a dimensionality of 1024. For the final edge energy MLP, a single hidden layer with a dimension of 128 was implemented without incorporating any nonlinearities. During training, a batch size of 5 and a learning rate of 0.002 were utilized to update the model parameters iteratively. The loss function of energies and forces to be minimized by the Allegro model architecture is presented in equation (1).

$$L = \frac{\lambda_E}{B}\sum_b^B\left(\hat{E}_b - E_b\right)^2 + \frac{\lambda_F}{3BN}\sum_{i=1}^{BN}\sum_{\alpha=1}^{3}\left\|-\frac{\partial \hat{E}}{\partial r_{i,\alpha}} - F_{i,\alpha}\right\|^2. \tag{1}$$

Where $B, N, E_b, \hat{E}_b, F_{i,\alpha}$ refer to the batch size, number of atoms, batch of reference energies, batch of predicted energies, and the force component on atom $i$ in the spatial direction $\alpha$, respectively and $\lambda_E, \lambda_F$ are the energy and force weights. The initial coefficients for the energy and forces in the loss function were set to 1.0. For more detailed information, visit the original publication for Allegro[67]. After training, the models were tested to predict the potential energy and forces of the configurations in the test data set.

### 2.5 Deep Potential Molecular Dynamics Simulations

Trained models were employed to conduct deep potential molecular dynamics (DPMD) utilizing LAMMPS coupled with Allegro and PLUMED. The systems for these simulations were identical to those constructed for the reference data, consisting of a single molecule enclosed within a cubic box with sides measuring 4 nm. The trained Allegro MLFFs were used to describe the interactions of atoms in the system and PLUMED was used to track the sampling of CVs. DPMD-biased simulations allowed the evaluation of the MLP's accuracy in predicting the FES of butane and ADP. Metadynamics simulations were conducted at 300 K for ten nanoseconds. For butane simulations, Gaussian curves were added every 500 steps with a height of 0.28 kcal/mol, a width of 0.3 rad, and a bias factor of 5. While ADP simulations were conducted at 300 K, Gaussian curves were added every 100 steps with a height of 0.3 kcal/mol, a width of 0.3 rad, and a bias factor of 3.

### 2.6 Data Availability

The GitHub repository of this project contains all the files required to generate training/test data, train the models, and use them for metadynamics simulations.

- *https://github.com/omendibleba/Considerations_for_MLPs_FES*

## 3 Results

We evaluate Allegro's accuracy in recovering the FES of butane and ADP using four distinct data sets at the classical and *ab initio* levels of theory. The datasets aim to replicate the sampling that could be obtained in hypothetical scenarios where the system's FES is not well studied. Additionally, we test the potential effect of fixing bond lengths and angles in the training datasets. The three tests are:

1. Learning butane's FES at the classical and *ab initio* level of theory.
2. Learning ADP's FES at the classical level of theory with and without bond/angle constraints.
3. Learning ADP's FES at the *ab initio* level of theory with and without bond/angle constraints.

## 3.1 Test 1

The first test aimed to study the effect of the CV distribution in training data on the accuracy of butane's free energy predictions. Figure 3 presents the predicted FES from the models trained at the classical and *ab initio* levels of accuracy in panels A and B, respectively.

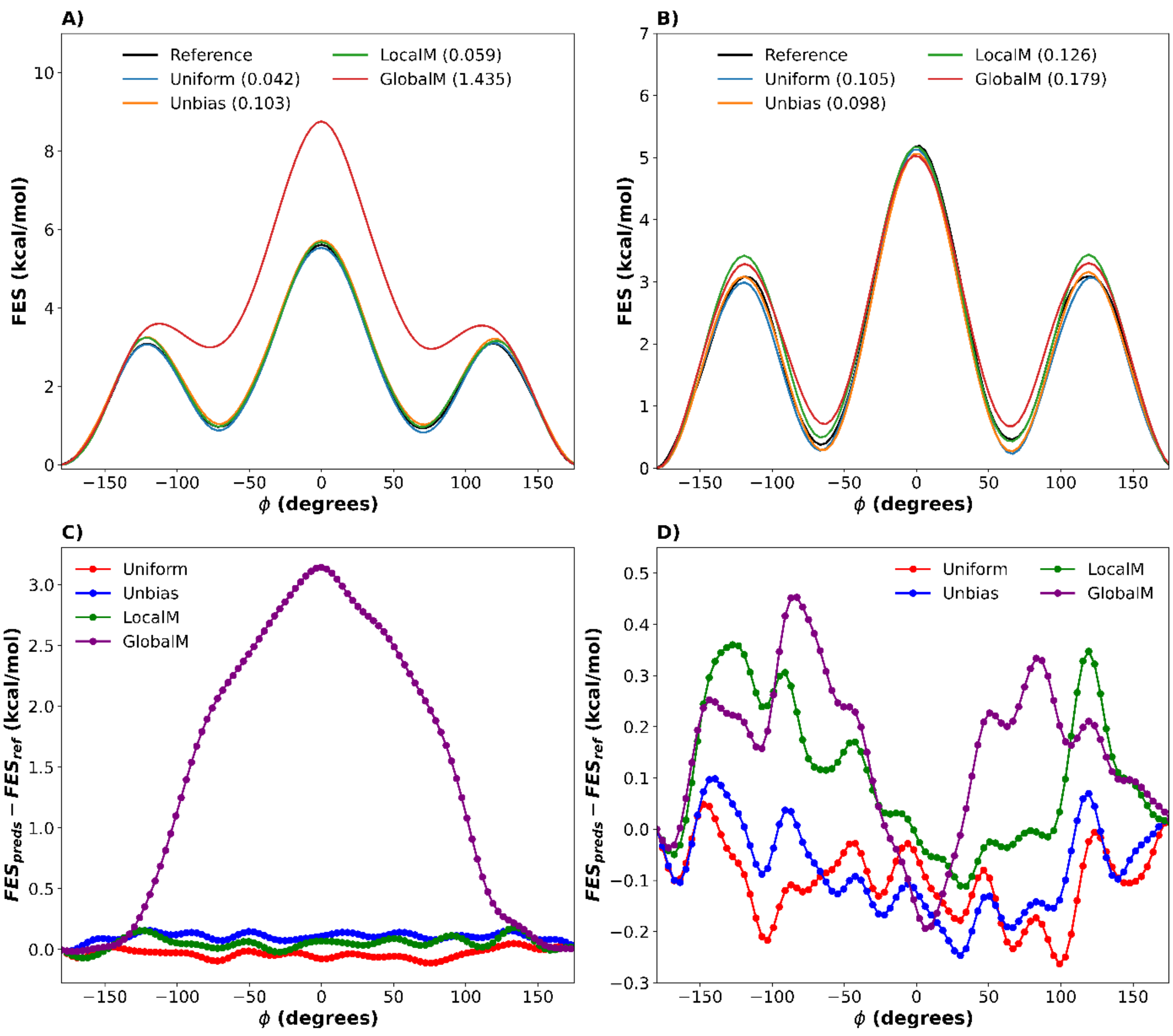

*Figure 3. Predicted free energy surface for Allegro models trained with classical and ab initio levels of theory in A) and B), respectively, compared to the reference FES at the same level of theory. Mean absolute error (MAE) values are presented in*

*parenthesis and kcal/mol. Plots C) and D) show the difference between the predicted and reference FES at the classical and Ab Initio levels of theory, respectively.*

Except for the Global Minimum model, the models trained with energies and forces from classical MD were able to recover the system's FES at the same level of theory. The model was accurate in regions where it was trained, and the error increased as the CV value of configuration was farther from those included in the training data. This model, however, was not accurate in extrapolating the free energy of configurations near 0° and resulted in only a qualitative recovery of the FES's shape. Interestingly, when the same configurations are used with energies and forces at the *ab initio* level of theory, the model can extrapolate the energies near 0° more accurately. This is evident by a reduction of the mean absolute error (MAE) between the predicted and reference FES from 1.44 kcal/mol to 0.179 kcal/mol when trained with classical versus *ab initio* data. The difference between the predicted and reference FES was determined to analyze further the lack of accuracy of the GlobalM model trained at the classical versus *ab initio* level of theory.

Figure 3 C highlights the increase in the error of the model trained with configurations only representing the global minima of the system's free energy at the classical theory level. The difference between the predicted and reference data near the dihedral angle at 0° reaches 3 kcal/mol and reduces to zero as the configurations reach the 180° degree angle, where training data is given. This is an expected result since the model should perform better in areas of the FES observed during training and lose accuracy as sampled configurations in MD are farther from it. Interestingly, when the same configurations are used, but their energies and atomic forces are determined with *ab initio* calculations, the difference between predicted and reference values is maintained under 0.5 kcal/mol, as shown in Figure 3 D.

These results indicate that, at the classical level of theory, the distribution of CVs did not significantly affect the accuracy of trained models if characteristic regions of the FES were represented by training data configurations. Additionally, independently of the training distribution used for butane models, if the energies and forces were obtained with *ab initio* calculations, the MLFFs could accurately recover the system's FES. Predictions made by the MLFFs trained with the LocalM and GlobalM distribution highlight Allegro's capability to extrapolate the energies and forces of configurations not included during training. Additionally, the accurate recovery of the FES from the distinct MLFFs serves as additional proof of concept for a general workflow for training MLFFs. One that involves the generation of configurations with cheap classical molecular dynamics, calculating their energies and forces with DFT or other *ab initio* approach, and using these data to train the MLFF. The recovery of the FES from the unbiased model implies that for small molecules like butane, the configurations obtained from unbiased simulations are sufficient to train a model capable of recovering the entire FES of the system with first-principle calculation accuracy.

## 3.2 Test 2

The following study focuses on the effect of the CV distribution in training data with and without fixed bonds and angles on the accuracy of ADP's free energy predictions at the classical level of theory.

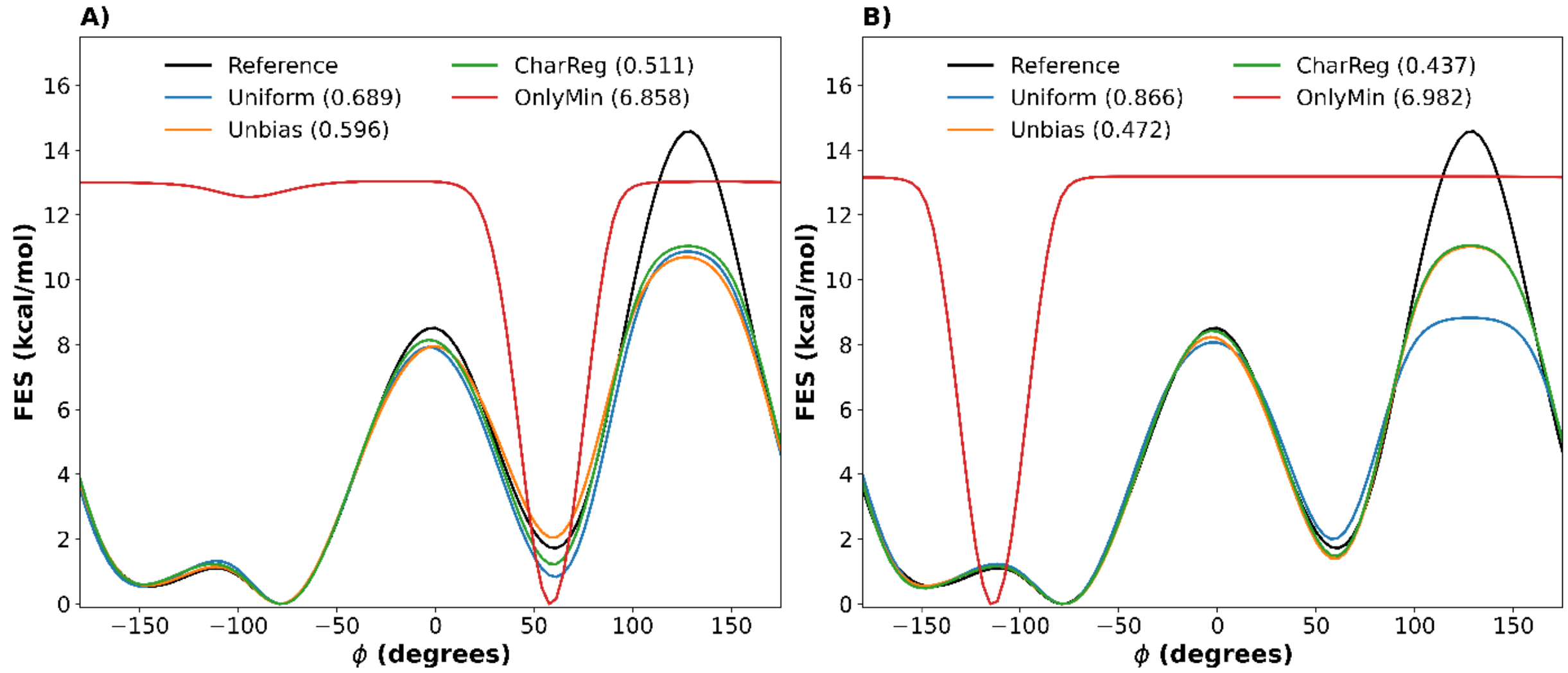

*Figure 4. Predicted free energy surface for Allegro models trained with a classical level of theory with and without bond-angle constraints in A) and B), respectively. Mean absolute error (MAE) values are presented in parenthesis and kcal/mol.*

The Allegro models trained for ADP at the classical level of theory failed to accurately recover the complete FES of the system independently of the addition or exclusion of angle and bond constraints (see Figure 4). Models trained with Uniform, Unbiased, and Characteristic Region configurations failed to predict the free energy of configurations with phi values near 120°, which is where the highest free energy barrier occurs. Error in this region is attributed to a lack of distinct configurations with similar CV values but distinct configurations of the atoms that do not form part of the CV.

Models trained with sampling only on the minimum of the FES severely failed to recover the FES of the system, as evidenced by MAE higher than 6 kcal/mol. A slight increase in variability is observed for the FES predictions of models trained with bond and angle constraints compared to those without them. This variability can be related to the model's training and not necessarily to the data. These results suggest that no significant effect is observed due to using bond and angle constraints when generating configurations using classical MD. Like Test 1’s results, the distribution of CVs does not significantly affect the MLFFs' accuracy in predicting the system's FES. However, the lack of configurations in high-free energy regions can result in a systematic accuracy error. In this context, the accuracy of classical simulations may not be sufficient to differentiate two very similar configurations energetically. Independent of the training distribution and the use of bond-angle constraints, the trained models were inaccurate in the region of higher free energy but could recover the overall shape of the FES.

## 3.3 Test 3

In the final test, we studied the effect of the CV distribution in training data with and without constraints on the accuracy of ADP’s free energy predictions at the *ab initio* level of accuracy.

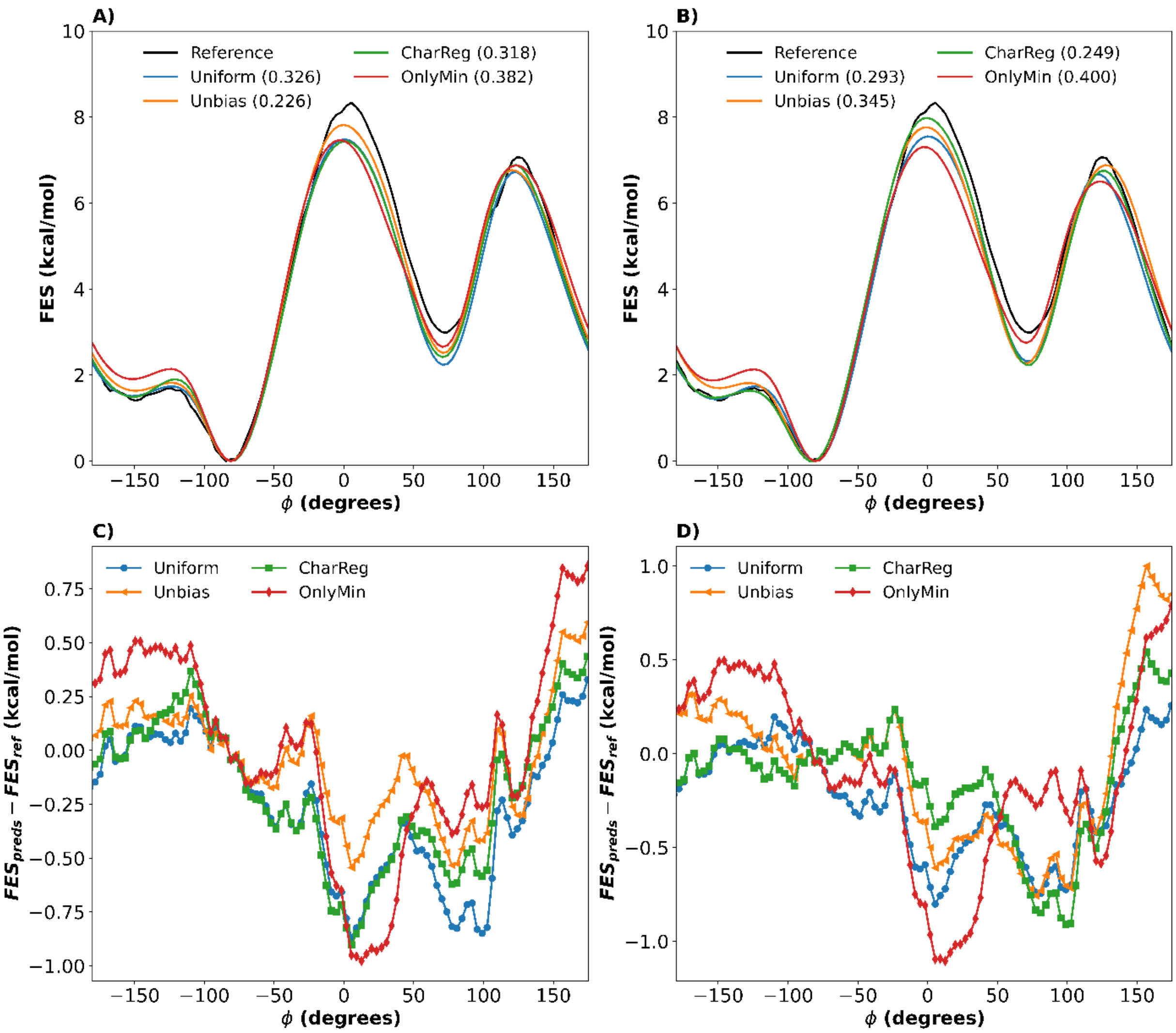

*Figure 5. Predicted free energy surface for Allegro models trained with an ab initio level of theory with and without constraints in A) and B), respectively. Mean absolute error (MAE) values are presented in parenthesis and kcal/mol. Plots C) and D) present the difference between the predicted and reference FES at the classical and Ab Initio levels of theory, respectively.*

We trained models for this test using the same configurations used in test four but with energies and forces from *ab initio* single-point calculations. The results in panels A and B of Figure 5 show that independently of the training distribution and usage of constraints, the models could recover the FES of ADP as a function of its phi dihedral at the *ab initio* level of theory. In contrast with the previous test, the training configurations appear to be sufficient to recover the system's FES. Therefore, the lack of accuracy in the previous test may be related to the level of accuracy of the calculation for energies and forces and not the actual configurations included in the training data. Energies and forces can have a smaller magnitude difference in classical MD than when using *ab initio* approaches. As a result, if a configuration is not included in the data set, extrapolations of the model become less accurate with the narrower distribution. Similar to the previous results, maintaining the length of bonds and angles of the configurations did not impact the accuracy of the FES predictions at this level of theory. Panels C and D in Figure 5 highlight similar trends in the difference between predicted and reference values, further evidencing the negligible effect of generating configurations using bond and angle constraints. These differences also evidence that the OnlyMin model had the largest deviations at both the high and low energy regions, but transition areas were captured accurately. In contrast, the model trained with the Uniform data presented more accuracy across the entire surface. This suggests that if the FES of the system of interest is known, generating training data that samples the whole surface can produce an MLFF capable of recovering it. For cases in which the FES is not known, the MLFFs can make accurate predictions, which accuracy scales with the complexity of the FES of interest. These results highlight the capability of Allegro models to recover the FES of a system even when the training data is missing important configurations that represent transition areas between free energy minima and maxima. Results from the Unbiased and OnlyMin models trained without constraints, with an MAE of 0.326 kcal/mol and 0.382 kcal/mol, respectively, further validate this conclusion. Moreover, the results validate the idea of the workflow for generating training data. Generating configurations with classical MD and then using DFT for single-point calculations provides an easy workflow for generating training data for virtually any system.

## 3.4 Lessons learned

This section includes important lessons learned during this investigation that will benefit future research using MLFF for the prediction of FES through enhanced sampling simulations.

1. Analyzing training and test data energy distributions

The importance of analyzing the distribution of potential energies included in the training and test set increases with the complexity of the system under study. Analyzing the distribution of CVs gives important information about what parts of the FES are represented in the training data, but if the FES of the system is unknown, we can use the distribution of PE to determine if sufficient configurations are included in the training data. We compare the potential energy distributions used for the ADP Uniform distribution model with and without constraints in Figure 6.

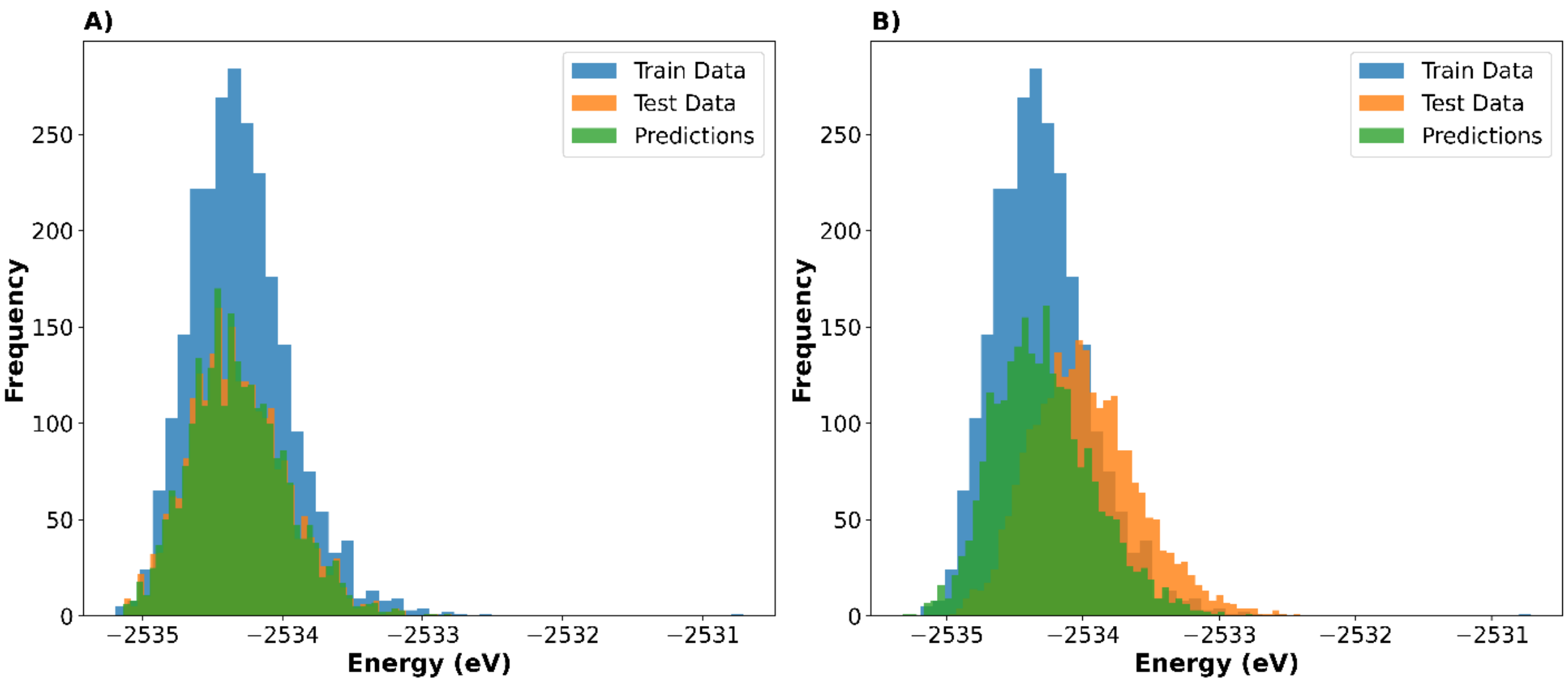

*Figure 6. Histogram of potential energies included in the AD Uniform distribution model in the training data, test data, and predictions on test data with and without constraints at ab initio level of theory in A) and B), respectively. Mean absolute error (MAE) values are presented in parenthesis and kcal/mol. Plots C) and D) present the difference between the predicted and reference FES at the classical and ab initio levels of theory, respectively.*

When the configurations are generated using bond and angle constraints (Figure 6 A), the configurations with the same CV values will have virtually the same configuration. In contrast, when the configurations are generated without constraints, configurations with the same CV might have slightly different configurations due to other atoms in the system that are not part of the CV, also moving and effectively changing the potential energy of the configurations and, therefore, its free energy. For this reason, if we use the same approach for generating training and test data, it is essential to ensure that the test data and training data have a similar distribution in potential energy. Also, the distributions do not match, which indicates that both data sets lack information, and the suggestion is to generate more configurations until both distributions match.

2. Potential energy predictions of a test data set do not reflect the real accuracy of the model

Ensuring the train and test data have matching potential energy distribution can have negative implications when analyzing the accuracy of the trained model using a parity plot that compares predicted vs. reference values. Panels A and B in Figure 7 present the parity plots obtained from evaluating trained models (with bond and angle constraints) against the test data set at the classical and ab initio levels of theory, respectively. When the configurations in the train and test data fall under similar distributions of potential energy, the model is expected to accurately predict the energies, which is what we see in our results, with the exception of the OnlyMin model trained with classical data. The MAE of the predicted values in this test is commonly reported as a general accuracy metric of the model, but this accuracy is not necessarily transferable for other tests, such as the prediction of the FES of the molecule. Panels C and D in Figure 7 show a parity plot for the predicted and reference-free energy values. The MAE of these tests is approximately five times larger than the MAE of potential energy predictions for the MLFFs trained with classical data and about 2.5 times for force fields trained with *ab initio* data. Therefore, the accuracy of the potential energy test can give some insight into which model will be more accurate in general, but a case-specific test must be conducted as well until a fully robust accuracy evaluation protocol is developed.

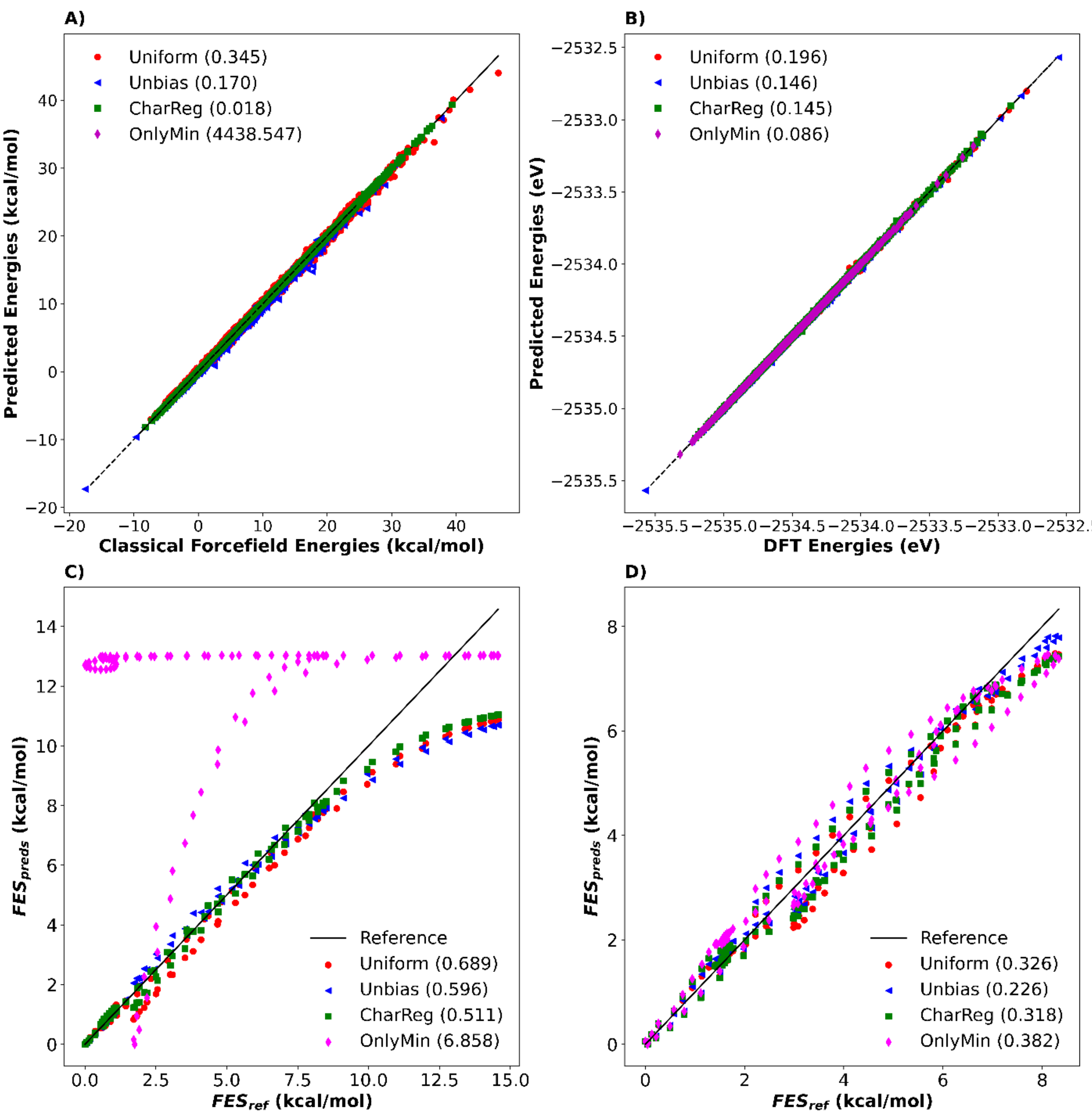

*Figure 7. Parity plot of predicted and reference potential energy values of alanine dipeptide from test dataset with constraints at the classical and ab initio levels of theory in A and B, respectively. Plots C and D show alanine dipeptide's predicted and reference free energy values with constraints at the classical and ab initio levels of theory. Mean absolute error (MAE) values are presented in parenthesis and the same units as the plot's axis.*

Depending on the application for which the model is being trained, this could be a problem. Suppose a model is being trained for structural analysis, for example, and the training and test configurations include all relevant configurations, which can be representative of local minima in free energy. In that case, the accuracy of this test might reflect the accuracy of the structural analysis test. If the same model is used to predict the FES of the system, then the model can show less accuracy. This analysis highlights again the importance of generating training and test configurations using different approaches to ensure that potential energy distributions converge to a similar shape and mean. This behavior seems more relevant when the training data comes from classical force fields than from first principle calculations. However, we consider it good practice to include all possible configurations in the training data of the EQNN, even if the purpose of the model is to be used with a subset of configurations.

3. When overfitting is not a problem

When training neural network models, it is helpful to visualize the training and validation loss functions as a function of training epochs. By comparing their values, one gets a notion of model overfitting or underfitting issues, which are usually remediated by adjusting the parameters of the model and training procedure. Interestingly, results from this work indicate that with this specific type of neural network architecture (i.e., EQNN) and for the studied systems, overfitting might be required for accurate FES predictions. Figure 8 A shows the learning and loss function of a normal model (N), an over-parametrized model (OP), and an over-parametrized model with controlled epochs (OP-C). The normal model has fewer parameters (8.1 million) than the one used for the result discussed in previous sections, which is the OP model (11.2 million), and its loss function did not improve with the number of epochs. The OP model was trained until the validation loss function was minimized and did not change for a hundred epochs, at this point, an early stopping function was activated. Since the training loss of this model was an order of magnitude larger than the validation, this model is considered overfitted. To test a not-overfitted model, we trained the OP-C model by strictly defining the number of epochs to train the model to 290. This value was selected based on the loss function of the OP model, which shows that after 290 epochs, the training loss function fell under the validation loss.

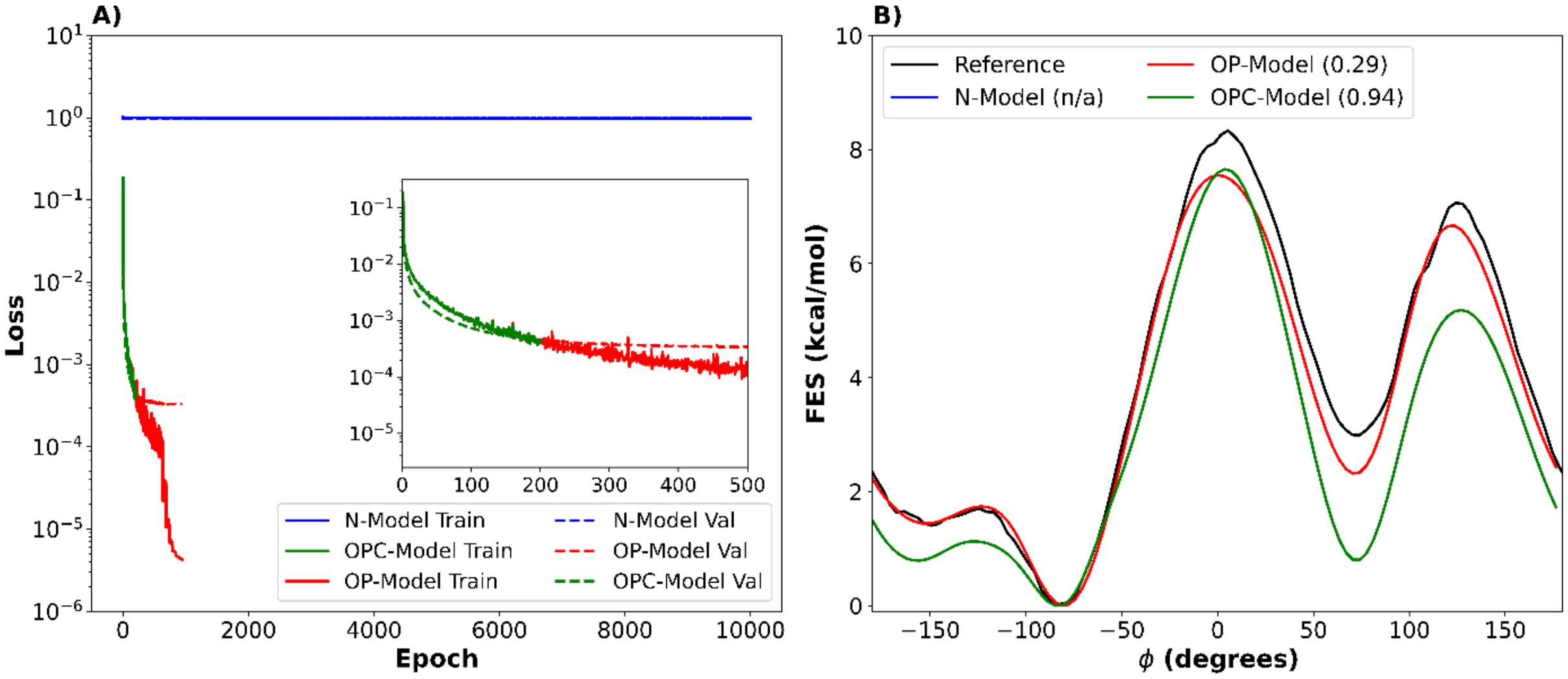

*Figure 8. Training and validation loss in A) and predicted FES in B) of the normal model (N), over-parametrized model (OP), and the over-parametrized model with controlled epochs (OP-C). All models were trained in uniform data distribution at the ab initio level of accuracy.*

The loss functions of the N-model indicate that the parameters were not properly optimized for the input data. This model's training and validation loss remained without minimization for ten thousand epochs with loss values greater than 10-1, suggesting that a more robust architecture could be needed to describe the data of interest. For this reason, we used the OP model, and the losses seemed to be adequately minimized for approximately 400 epochs, achieving loss values under 10-3. After approximately 500 epochs, the training loss decreases rapidly while the validation loss is maintained at the same level, causing a behavior interpreted as an overfitted model. We tried various hyperparameter combinations to mitigate the overfitting of the model, but the results were similar to those observed for the normal model. For this reason, we decided to retrain the model with the same hyperparameters but stop the training before overfitting behavior was observed and train the OP-C models. The loss functions of this model showed that the training stopped at around 250 epochs, and the values fell under 10-3. These models were then tested for their ability to recover the FES of ADP as a function of its phi dihedral, and the results are presented in Figure 8 B. The normal model failed to adequately describe the interactions in the molecule and presented unphysical configurations during the simulation run, destabilizing the system and making it crash, finally failing to recover the FES. The OP models could recover the system's FES with different levels of accuracy. The OP model was the most accurate overall, and its regions of higher error are located at the free energy peaks at 0° and 120° and the local minimum near 60°. On the contrary, the OPC model was able to capture the CV values at which the peaks and minima occur but underpredicted the system's FES, except the global minimum, which was accurately predicted. These results indicate the overfitted model was better at recovering the FES of the system than other models. Predictions of overfitted models in this sense are attributed to the system's relative simplicity. For ADP, the distinct configurations for local environments around each atom are fewer than what can be expected on solvated or solid systems. For this reason, if the training of the local environments in the training data includes all the local environments that can be observed during the metadynamics simulation, the model will not need to be extrapolated and will be perceived as accurate. However, suppose the system is more complex and not all local atomic environments are accounted for in the training configurations. In that case, the behavior observed in this test is not necessarily expected. Additional research is required to obtain more information about the usability of overfitted MLFFs for FES predictions at the ab initio level of theory.

# 4 Conclusions

This study examines the effect that sampled CVs in training data have on the accuracy of EQNN in predicting the FES of butane and ADP at the classical and ab initio levels of theory. This was achieved by generating four training data sets for butane and three for ADP, each with different CV distributions. The workflow used to create configurations of interest represents a new methodology to generate configurations representing regions of high free energy, which could be hard to obtain otherwise. Moreover, the methodology involves the generation of coordinates with low-cost molecular dynamics and the calculation of energies and forces with SPCs at the ab initio level of theory. This reduces the cost involved in moving the atoms forward in time using AIMD directly and allows for multiple SPCs to run in parallel. FES predictions of butane models indicate that for such a simple system, the distribution of CVs in the training data does not play a significant role if all the regions of the configurational space are represented. These predictions highlight the extrapolation capabilities of the Allegro models to accurately predict the free energy of configurations not included in the training data. Models trained from classical MD data showed similar MAEs on free energy predictions when the training data included and excluded bond and angle constraints. This result indicates that the inclusion of bond and angle constraints in MD simulations did not significantly affect the accuracy of the models' free energy prediction. Results from ADP models trained with ab initio data could reproduce the FES of ADP as a function of its phi dihedral at the same level of theory. This result validated the usability of the developed workflow to train EQNNs capable of predicting the FES of a system even when not all the configurations that represent the FES's maxima and minima are available in the training data. In addition to these results, three important takeaway lessons are discussed. First, it is important to analyze and share histograms of potential energy since they can provide helpful information about the robustness of the training data and can have significant implications on the results of energy prediction tests. Second, the accuracy of the potential energy prediction test does not necessarily reflect on the accuracy of FES prediction through metadynamics simulations. For this reason, we consider that predictions of the FES of the system, when available, is a better accuracy metric to report. Finally, we highlight the capability of overfitted models to recover FES accurately and the lack of accuracy of not overfitted models in doing so. We believe that the knowledge we have attained will help train EQNNs on more complex systems and achieve accurate predictions of FESs of systems that are currently unattainable.

## Supplementary material

The supplementary material includes convergence results for the reference and predicted FES of butane and ADP. In addition, it contains information about hyperparameter optimization for the used EQNNs.

## Acknowledgements

This work was performed using the computational resources provided by the Notre Dame Center for Research Computing (NDCRC). Y.C. acknowledges funding support from the U.S. Department of Education (Award No. P200A210048) and the University of Notre Dame. J.K.W. acknowledges support from the Department of Energy, Basic Energy Sciences, Materials Science, and Engineering Division, through the Midwest Integrated Center for Computational Materials (MICCoM).

## Author Declarations

### Conflict of Interest

The authors have no conflicts to disclose. Author Contributions Orlando A. Mendible-Barreto: Data curation (lead); Formal analysis (equal); Investigation (lead); Methodology (equal); Visualization (lead); Writing – original draft (lead); Writing – review & editing (lead). Jonathan K. Whitmer: Conceptualization (equal); Project administration (supporting); Writing – review & editing (supporting). Yamil J. Colón: Conceptualization (lead); Funding acquisition (lead); Methodology (equal); Project administration (lead); Supervision (equal); Writing – review & editing (supporting).

## Data Availability

The GitHub repository of this project contains all the files required to generate training/test data, train the models, and use them for metadynamics simulations. https://github.com /omendibleba/Considerations_for_MLPs_FES.